\documentstyle[12pt]{article}

\input{tcilatex}
\begin{document}

\title{Single gluino production in the R-parity lepton number violating MSSM at the LHC
\footnote{The project supported by National Natural Science
          Foundation of China}}
\author{{\ Yin Xi$^{b}$, Ma Wen-Gan$^{a,b}$, Wan Lang-Hui$^{b}$ and Jiang Yi$^{b}$}
\\
{\small $^{a}$CCAST (World Laboratory), P.O.Box 8730, Beijing 100080, China}
\\
{\small $^{b}$Modern Physics Department, University of Science and}\\
{\small Technology of China, Hefei, Anhui 230027, China.}}
\date{}
\maketitle

\begin{abstract}
We examine the $R_{p}$-violating signal of single gluino
production associated with a charged lepton or neutrino at the
large hadron collider (LHC), in the model of R-parity relaxed
supersymmetric model. If the parameters in the ${\rlap/R}_p$
supersymmetric interactions are not too small, and the mass of
gluino is considered in the range from several GeV (as the
Lightest Supersymmetric Particle) to 800 GeV, the cross section of
the single gluino production via Drell-Yan processes can be in the
order of $10^2 \sim 10^3$ femto barn, and that via gluon fusion in
the order of $10^{-1} \sim 10^3$ femto barn. If the gluino decay
can be well detected in the CERN LHC, this process provides a
prospective way to probe supersymmetry and $R_p$ violation.
\newline
\end{abstract}



\vspace{8mm}


\vfill \eject

\baselineskip=0.36in


\noindent{\large {\bf I. Introduction}}
\par
The new physics beyond the standard model (SM) has been
intensively studied in the past years\cite{beyondSM}. The
supersymmetric models (SUSY) are the most attractive ones among
the general extended models of the SM. As we know that electroweak
gauge invariance requires absence of the terms in the lagrangian
that change either baryon number or lepton number. Usually these
terms may lead to unacceptable short proton lifetime. One way to
solve the proton-decay problem is to impose a discrete symmetry
conservation called R-parity ($R_p$) conservation\cite{rdef}.
Actually this conservation is put into the MSSM for the purpose to
retain the symmetries of the SM. But the most general SUSY
extension of the SM should contain such terms.
\par
If the R-parity is conserved, all supersymmetric partner particles
must be produced in pair, thus the lightest of superparticles must
be stable. If R-parity is violated, the feature of the SUSY
particles changes a lot. Until now we have been lacking in
credible theoretical argument and experimental tests for $R_p$
conservation, we can say that the $R_p$ violation
(${\rlap/{R}}_p$) would be equally well motivated in the
supersymmetric extension of the SM. Finding the signature of $R_p$
violation has recently motivated some investigation\cite{signals}
because of experimentally observed discrepancies.
\par
Experimentally searching for the effects of ${\rlap/{R}}_p$
interactions has been done with many efforts in the last few
years. Unfortunately, up to now we have only some upper limits on
${\rlap/{R}}_p$ parameters. It is necessary to continue these
works on finding ${\rlap/{R}}_p$ signal or getting further
stringent constraints on the ${\rlap/{R}}_p$ parameters in future
experiments. Detecting signals of the supersymmetric particle is a
prospective way in searching for R-parity violation. The process
of single chargino/neutralino production with ${\rlap/R}_p$ has
been intensively studied in former works\cite{wanfeng}. The
strongly interacting supersymmetric paticles, squark $\tilde{q}$
and gluino $\tilde{g}$, can be produced with the largest cross
sections at hadron colliders. From the evolution of parameters,
$\tilde{g}$ should be the heaviest gauginos at low scale, since
the ratios of gaugino masses to coupling constants do not change
with scale in one-loop approximation\cite{mSUGRA}\cite{heavyg}.
However, in some other models gluino may exist as the Lightest
Supersymmetric Particle (LSP)\cite{gLSP}. The production and decay
of gluino with or without R-parity conservation have been
investigated in \cite{gdecay} \cite{lambda}. It is showed in
Ref.\cite{rgdecay} that if gluino is lighter than $\sim 1 $ TeV,
the signal of $\tilde{g}$ decay can be detected by initial
searches for sparticles at the LHC even if R-parity is not
conserved. In Ref.\cite{yu} single gluino production via Drell-Yan
process with R-parity baryon number violation has been studied. It
is showed that the production of gluino can be well detected when
the ${\rlap/{R}}_p$ parameters are not too small, e.g.
$\lambda^{\prime \prime}=0.01\sim 0.1$.
\par
In this paper we studied the single gluino production associated
with a charged lepton or neutrino in the framework of the MSSM
with R-parity lepton number violation for both the Drell-Yan
process and the one-loop gluon fusion process. In section II we
present the model and calculation of the process $pp \rightarrow
\tilde{g}+X$. Numerical results and discussion are given in
section III. In section IV we give a short summary. The details of
expressions in the calculation can be found in the Appendix.

\begin{flushleft} {\bf II. Calculation} \end{flushleft}
\par
The general ${\rlap/R}_p$ superpotential can be written as
\begin{equation}
W_{\rlap/{R}_p} = \epsilon_{ij} (\lambda_{IJK} \tilde{L}_i^I
\tilde{L}_j^J \tilde{R}^K+\lambda_{IJK}^{\prime}\tilde{L}_i^I
\tilde{Q}_j^J \tilde{D}^K+ \epsilon_I H_i^2
\tilde{L}_j^I)+\lambda_{IJK}^{\prime\prime}\tilde{U}^I \tilde{D}^J
\tilde{D}^K
\end{equation}

$L^I, Q^I, H^I$ denotes the SU(2) doublets of lepton, quark and
Higgs superfields respectively, while $R^I, U^I, D^I$ are the
singlets lepton and quark superfields. The bilinear
${\rlap/{R}}_p$ term $\epsilon_{i j} \epsilon_I H_i^2
\tilde{L}_j^I$ will lead to the mixture of mass eigenstates and
give the neutrino masses. However, their effects are assumed to be
negligible in our process. The constraints on the
couplings\cite{lambda},
\begin{equation}
\mid(\lambda~or~\lambda^{\prime}) \lambda^{\prime\prime}\mid<10^{-10} \left(%
\frac{\tilde{m}}{100 GeV}\right)^2
\end{equation}
is usually taken to indicate that only lepton or baryon number violation
exist. We consider only the lepton number violation, i.e., $\lambda$ and $%
\lambda^{\prime}$ are assumed to be non-zero while $\lambda^{\prime\prime}$
terms are neglected.
\par
In this paper we denote the Drell-Yan tree-level processes
\begin{equation}
pp\rightarrow
u\bar{d}(\bar{u}d)\rightarrow\tilde{g}\tau^{+}(\tilde{g}\tau^{-})
\end{equation}
\begin{equation}
pp\rightarrow d\bar{d}\rightarrow\tilde{g}\nu_{\tau}
\end{equation}
and the one-loop process through gluon fusion
\begin{equation}
pp\rightarrow gg\rightarrow\tilde{g}\nu_{\tau}
\end{equation}
Because of charge conjugation invariance, the cross section of
subprocess $u\bar{d}\rightarrow\tilde{g}\tau^{+}$ coincides with
$\bar{u}d\rightarrow\tilde{g}\tau^{-}$. Then we give only the
calculations for the subprocess
$u\bar{d}\rightarrow\tilde{g}\tau^{+}$. We should mention that
there in no ${\rlap/R}_p$ coupling of $u\bar{\tilde{u}}\nu$
in the lowest order, so we don't consider the $u\bar{u}\rightarrow\tilde{g}%
\nu$ channel. The cross section of the Drell-Yan process are determined by ${%
\rlap/R}_p$ parameters of the first generation, i.e., $\lambda^{%
\prime}_{311} $, while the gluon fusion process depend on
${\rlap/R}_p$ parameters of all generations, especially
$\lambda^{\prime}_{333}$. Although in our calculation, the
contribution of the one-loop gluon fusion process appears to be
relatively small, they are not negligible due to the facts that
there is large gluon luminosity in protons and the ${\rlap/R}_p$
couplings of the third generation could be large.
\par
The Feynman diagrams for the tree-level process
$u\bar{d}\rightarrow\tilde{g} \tau$ and
$d\bar{d}\rightarrow\tilde{g}\nu_{\tau}$ are given in Fig.1 and
Fig.2 respectively. The $gg \rightarrow \tilde{g}\nu_{\tau}$
one-loop diagrams at the lowest order are plotted in Fig.3. In our
calculation the ultraviolet divergence in subprocess $gg
\rightarrow \tilde{g}\nu_{\tau}$ should be cancelled automatically
and it's not necessary to consider the renormalization at one-loop
level. Fig.3(a) contains the s-channel diagrams, Fig.3(b) the box
diagrams, and Fig.3(c) the quartic diagrams. The relevant Feynman
rules without $\rlap/R_p$ interactions can be found in
references\cite{rules}. The related Feynman rules with $\rlap/R_p$
interactions can be read out from Eq.(1), which are listed in the
Appendix.
\par
We define the Mandelstam variables as usual
\begin{equation}
\begin{array}{lll}
\hat{s} & = & (p_1+p_2)^2 = (k_1+k_2)^2 \\
\hat{t} & = & (p_1-k_1)^2 = (p_2-k_2)^2 \\
\hat{u} & = & (p_1-k_2)^2 = (p_2-k_1)^2
\end{array}
\end{equation}
The amplitude of $u\bar{d}\rightarrow\tilde{g}\tau^+$ (Feynman diagrams in
Fig.1) is given by
\begin{equation}
M_{u\bar{d}} = M_{u\bar{d}}^{\hat{t}} + M_{u\bar{d}}^{\hat{u}}
\end{equation}
where
\begin{equation}
\begin{array}{lll}
M_{u\bar{d}}^{\hat{t}} & = & \bar{u}(k_1)[i\sqrt{2} g_s
T_{\beta\alpha}^{\gamma}
(-Z_U^{*1i}P_L+Z_U^{*2i}P_R)]u(p_1)\frac{i} {\hat{t}-m_{\tilde{u}_i}^2}\bar{v}%
(p_2)(i\lambda^{\prime}_{311}Z_U^{1i}P_R) v(k_2) \\
M_{u\bar{d}}^{\hat{u}} & = & -\bar{v}(p_2)[i\sqrt{2} g_s
T_{\beta\alpha}^{\gamma}
(-Z_D^{*1i}P_R+Z_D^{*2i}P_L)]v(k_1)\frac{i} {\hat{u}-m_{\tilde{d}_i}^2}\bar{u}%
(k_2)(i\lambda^{\prime}_{311}Z_D^{*2i} P_L)u(p_1)
\end{array}
\end{equation}
$Z_D^{ij}$ and $Z_U^{ij}$ denote the matrices used to diagonalize
the down-type squark and up-type squark mass matrices,
respectively. $\alpha$, $\beta$ and $\gamma$ denote the color
indices of the initial up-quark, down-quark and final gluino
respectively. Similarly the amplitude of
$u\bar{d}\rightarrow\tilde{g}\nu_{\tau}$ (Feynman diagrams in
Fig.1) is given by
\begin{equation}
M_{d\bar{d}} = M_{d\bar{d}}^{\hat{t}} + M_{d\bar{d}}^{\hat{u}},
\end{equation}
where
\begin{equation}
\begin{array}{lll}
M_{d\bar{d}}^{\hat{t}} & = & \bar{u}(k_1)[i\sqrt{2} g_s
T_{\alpha\beta}^{\gamma}
(-Z_D^{1i}P_L+Z_D^{2i}P_R)]u(p_1)\frac{i} {\hat{t}-m_{\tilde{d}_i}^2} \\
&  & \times\bar{v}(p_2)[i\lambda^{\prime}_{311}(Z_D^{2i}P_R
+Z_D^{*1i}P_L)]v(k_2), \\
M_{d\bar{d}}^{\hat{u}} & = & -\bar{v}(p_2)[i\sqrt{2} g_s
T_{\beta\alpha}^{\gamma}
(-Z_D^{*1i}P_R+Z_D^{*2i}P_L)]v(k_1)\frac{i} {\hat{u}-m_{\tilde{d}_i}^2} \\
&  & \times\bar{u}(k_2)[i\lambda^{\prime}_{311}(Z_D^{*2i}P_L
+Z_D^{1i}P_R)]u(p_1),
\end{array}
\end{equation}
$\alpha$, $\beta$ and $\gamma$ denote the initial down-quarks and
final gluino respectively.
\par
The amplitudes squared summed over the spins and colors can be
written explicitly as follows, where we assume that $Z_D^{ij}$ and
$Z_U^{ij}$ are real.
\begin{equation}
\begin{array}{lll}
\sum\limits_{}^{}|M_{ud}|^2 & = & 8 g_s^2 (\lambda^{\prime}_{311})^2 \{ \frac{(Z_U^{1i})^2}{%
(\hat{t}-m_{\tilde{u}_i}^2)^2} (\hat{t}-m_{\tau}^2)(\hat{t}-m_{\tilde{g}}^2) +\frac{%
(Z_D^{2i})^2}{(\hat{u}-m_{\tilde{d}_i}^2)^2} (\hat{u}-m_{\tau}^2)(\hat{u}-m_{\tilde{g}}^2) \\
& - & \frac{(Z_U^{1i})^2 Z_D^{1j} Z_D^{2j}}{(\hat{t}-m_{\tilde{u}_i}^2) (\hat{u}-m_{%
\tilde{d}_j}^2)} [(\hat{t}-m_{\tau}^2)(\hat{t}-m_{\tilde{g}}^2) +
(\hat{u}-m_{\tau}^2)
(\hat{u}-m_{%
\tilde{g}}^2) - \hat{s} (\hat{s}-m_{\tilde{g}}^2-m_{\tau}^2) ] \} \\
\sum\limits_{}^{}|M_{dd}|^2 & = & 8 g_s^2 (\lambda^{\prime}_{311})^2 \{ \frac{1}{(\hat{t}-m_{\tilde{%
d}_i}^2)^2} \hat{t} (\hat{t}-m_{\tilde{g}}^2)
+\frac{1}{(\hat{u}-m_{\tilde{d}_i}^2)^2}
 \hat{u} (\hat{u}-m_{%
\tilde{g}}^2) \\
& - & \frac{2 Z_D^{1i} Z_D^{1j} Z_D^{2i}
Z_D^{2j}}{(\hat{t}-m_{\tilde{d}_i}^2) (\hat{u}-m_{\tilde{d}_j}^2)}
[\hat{t} (\hat{t}-m_{\tilde{g}}^2) + \hat{u}
(\hat{u}-m_{\tilde{g}}^2) - \hat{s} (\hat{s}-m_{\tilde{g}}^2) ] \}
\end{array}
\end{equation}
\par
The corresponding amplitude of $g(p_1,a,\mu )g(p_2,b,\nu )
\rightarrow \tilde{g}(k_1,c)\nu _\tau (k_2)$ (Feynman diagrams in
Fig.3) can be written as
\begin{equation}
\begin{array}{lll}
{\cal M} & = & {\cal M}^b + {\cal M}^q + {\cal M}^{tr} \\
& = & \epsilon^{\mu}(p_1)\epsilon^{\nu}(p_2)\bar{u}(k_1) \left\{
f_1 g_{\mu\nu}+ f_2 k_{1\mu}k_{1\nu}+ f_3 g_{\mu\nu}\gamma_5+ f_4
k_{1\mu}k_{1\nu}\gamma_5+ f_5 k_{1\nu}\gamma_{\mu}+
f_6 k_{1\mu}\gamma_{\nu} \right. \\
& + & f_7 g_{\mu\nu}{\rlap/ p}_1+ f_8 k_{1\mu}k_{1\nu}{\rlap/ p}_1+ f_9
g_{\mu\nu}{\rlap/ p}_2+ f_{10} k_{1\mu}k_{1\nu}{\rlap/ p}_2+ f_{11}
k_{1\nu}\gamma_5\gamma_{\mu}+ f_{12} k_{1\mu}\gamma_5\gamma_{\nu} \\
& + & f_{13} g_{\mu\nu}\gamma_5{\rlap/ p}_1+ f_{14} k_{1\mu}k_{1\nu}\gamma_5{%
\rlap/ p}_1+ f_{15} g_{\mu\nu}\gamma_5{\rlap/ p}_2+ f_{16}
k_{1\mu}k_{1\nu}\gamma_5{\rlap/ p}_2+ f_{17} \gamma_{\mu}\gamma_{\nu} \\
& + & f_{18} k_{1\nu}\gamma_{\mu}{\rlap/ p}_1+ f_{19} k_{1\mu}\gamma_{\nu}{%
\rlap/ p}_2+ f_{20} \gamma_5\gamma_{\mu}\gamma_{\nu}+ f_{21}
k_{1\nu}\gamma_5\gamma_{\mu}{\rlap/ p}_1+ f_{22} k_{1\mu}\gamma_5\gamma_{\nu}%
{\rlap/ p}_2 \\
& + & f_{23} \gamma_{\mu}\gamma_{\nu}{\rlap/ p}_1+ f_{24}
\gamma_{\mu}\gamma_{\nu}{\rlap/ p}_2+ f_{25} k_{1\nu}\gamma_{\mu}{\rlap/ p}_1%
{\rlap/ p}_2+ f_{26} k_{1\mu}\gamma_{\nu}{\rlap/ p}_1{\rlap/ p}_2+ f_{27}
\gamma_5\gamma_{\mu}\gamma_{\nu}{\rlap/ p}_1 \\
& + & f_{28} \gamma_5\gamma_{\mu}\gamma_{\nu}{\rlap/ p}_2+ f_{29}
k_{1\nu}\gamma_5\gamma_{\mu}{\rlap/ p}_1{\rlap/ p}_2+ f_{30}
k_{1\mu}\gamma_5\gamma_{\nu}{\rlap/ p}_1{\rlap/ p}_2+ f_{31}
\gamma_{\mu}\gamma_{\nu}{\rlap/ p}_1{\rlap/ p}_2 \\
& + & \left. f_{32} \gamma_5\gamma_{\mu}\gamma_{\nu}{\rlap/ p}_1{\rlap/ p}%
_2+ f_{33} g_{\mu\nu}{\rlap/ p}_1{\rlap/ p}_2+ f_{34} g_{\mu\nu}\gamma_5{%
\rlap/ p}_1{\rlap/ p}_2 \right\} v(k_2)
\end{array}
\end{equation}
where ${\cal M}^b, {\cal M}^q$ and ${\cal M}^{tr}$ are the matrix
elements contributed by box, quartic and triangle interaction
diagrams, respectively.
\par
The cross sections for the subprocesses
$u\bar{d}(\bar{u}d,d\bar{d})\rightarrow \tilde{g}\tau
^{+}(\tilde{g}\tau ^{-},\tilde{g}\nu _\tau )$  and $gg\rightarrow
\tilde{g}\nu _\tau $ can be obtained by using the following
equation
\begin{equation}
\hat{\sigma}(\hat{s})=\frac 1{16\pi \hat{s}^2}\int_{\hat{t}^{-}}^{\hat{t%
}^{+}}d\hat{t}~\bar{\sum\limits_{}^{}}|{\cal M}|^2.
\end{equation}
where the bar over sum means average over the initial spin and
color. In above equations, $\hat{t}$ is the momentum transfer
squared from one of the incoming particle to the gluino in the
final state. For the subprocesses $gg(d\bar{d})\rightarrow
\tilde{g}\nu _\tau$ we have
\[ \hat{t}^{\pm }=\frac 12\left[ (m_{\tilde{g}}^2+m_{\nu
_\tau }^2-\hat{s})\pm \sqrt{(m_{\tilde{g}}^2+m_{\nu _\tau
}^2-\hat{s})^2-4m_{\tilde{g}}^2m_{\nu _\tau }^2}\right] .
\]
and for the subprocesses $u\bar{d}(\bar{u}d)\rightarrow
\tilde{g}\tau ^{+}(\tilde{g}\tau ^{-})$,
\[ \hat{t}^{\pm }=\frac 12\left[ (m_{\tilde{g}}^2+m_{\tau }^2-
\hat{s})\pm \sqrt{(m_{\tilde{g}}^2+m_{\tau
}^2-\hat{s})^2-4m_{\tilde{g}}^2m_{\tau }^2}\right] .
\]
\par
With the results from Eq.(13), we can easily obtain the total
cross section at $pp$ collider by folding the cross section of
subprocess with the quark and gluon luminosity correspondingly.

\begin{equation}
\sigma(s)= \int_{(m_{\tilde{g}}+m_{\tau,\nu_{\tau}})^2/ s} ^{1} d \tau \frac{d%
{\cal L}_{ij}}{d \tau} \hat{\sigma}_{ij}(\hat{s}=\tau s),
\end{equation}

where $\sqrt{s}$ and $\sqrt{\hat{s}}$ are the $pp$ collision and subprocess
c.m.s. energies respectively and $d{\cal L}_{ij}/d \tau$ is the distribution
function of parton luminosity, which is defined as
\begin{equation}
\frac{d{\cal L}_{ij}}{d\tau}=\frac{1}{1+\delta_{ij}}\int_{\tau}^{1} \frac{%
dx_1}{x_1} \left\{\left[ f_i(x_1,Q^2)f_j(\frac{\tau}{x_1},Q^2)
\right] +\left[ f_j(x_1,Q^2)f_i(\frac{\tau}{x_1},Q^2)
\right]\right\},~~~~ (i\geq j)
\end{equation}
here $\tau = x_1~x_2$, the definition of $x_1$ and $x_2$ are from
Ref.\cite{x1x2}, and in our calculation we adopt the MRS set G
parton distribution function \cite{function}. $f_{i,j}(x_n, Q^2)$
are the coreesponding quark and gluon distribution functions of
protons. The factorization scale Q was chosen as the average of
the final particles masses $\frac{1}{2}(m_{\tilde{g}} +
m_{\tau,\nu_{\tau}})$.

\vskip 5mm \noindent{\large {\bf III. Numerical Calculations and
Discussions}}
\par
In our numerical calculation to get the low energy scenario from
the MSUGRA \cite{mSUGRA}, the renormalization group
equations(RGE's)\cite{RGE} are run from the weak scale $m_Z$ up to
the GUT scale, taking all thresholds into account. We use two loop
RGE's only for the gauge couplings and the one-loop RGE's for the
other supersymmetric parameters. The GUT scale boundary conditions
are imposed and the RGE's are run back to $m_Z$, again taking
threshold into account. The effects of ${\rlap/R}_p$ to RGE's are
assumed to be small, and ${\rlap/R}_p$ parameters in the weak
scale are directly taken under the experimental upper bounds.
\par
Since we consider this process via lepton number violation terms
in the ${\rlap/R}_p$ model, and the ${\rlap/R}_p$ couplings
$\lambda^{\prime}$ in the terms of Eq.(1) inducing heavy lepton
can be very large from present upper limits\cite{limits}, we
choose $\lambda^{\prime}_{311}=0.05, \lambda^{\prime}_{322}=0.18$
and $\lambda^{\prime}_{333}=0.39$, and the other trilinear
parameters, i.e., $ \lambda$, $\lambda^{\prime}_{1ij}$,
$\lambda^{\prime}_{2ij}$  and $\lambda^{\prime\prime}$ have no
contribution to our process. The effect of bilinear breaking terms
are assumed to be very small and can be negligible. The SM input
parameters\cite{inputpara} are chosen as: $m_t=173.8~GeV,
m_Z=91.187~GeV, m_b=4.5~GeV, \sin^2\theta_W=0.2315$, and
$\alpha_{EW}=1/128$. We take a simple one-loop formula for the
running strong coupling constant $\alpha_s$
\begin{equation}
\alpha_s(\mu)=\frac{\alpha_s(m_Z)}{1+\frac{33-2n_f}{6\pi}\alpha_s(m_Z)
\ln \frac{\mu}{m_Z}}
\end{equation}
where $\alpha_s(m_Z)=0.117$ and $n_f$ is the number of active flavors at the
energy scale $\mu$.
\par
If the single gluino is produced in association with a charged
lepton, here assumed to be $\tau$, both final particles can be
well detectable. When a neutrino is produced instead of a charged
lepton, it would lead to an energy missing. If the gluino decay
could be well detected by the detector, it provides a prospective
way to observe the ${\rlap/R}_p$ process. For a heavy gluino, the
decay channel
\begin{equation}
\tilde{g}\rightarrow q\tilde{q}
\end{equation}
will dominate if kinematically
allowed. Since the $R_p$ violation parameters are strongly
constrained by experimental results, then the dominant subsequent
decays of squark are
\begin{equation}
\tilde{q}\rightarrow q\tilde{\chi}_i^0, q^{\prime}\tilde{\chi}%
_i^{\pm}
\end{equation}
where $\tilde{\chi}^0_i$ (i=1-4) denote the neutralinos and
$\tilde{\chi}^{\pm}_j$ (j=1,2) the charginos. The charginos and
neutralinos may decay further as
\begin{equation}
\tilde{\chi}^{\pm}_j \rightarrow \tilde{\chi}^0_{i} q
\bar{q'},(i=1-4,~j=1,2), ~~~~~~ \tilde{\chi}^{0}_k \rightarrow
\tilde{\chi}^0_{1} q \bar{q},(k=2,3,4)
\end{equation}
Therefore, like the R-parity conservation consequence, the typical
signatures for gluinos would mainly be two, four or six jets and
missing energy, carried away by the possible LSP
$\tilde{\chi}^0_{1}$. If squarks are heavier than gluino, the
following decays are possible:
\begin{equation}
\tilde{g}\rightarrow q\bar{q}\tilde{\chi}_i^0,
~~q\bar{q}^{\prime}\tilde{\chi}_i^{\pm}, ~~g\tilde{\chi}_i^0
\end{equation}
The following decays may be important for large ${\rlap/R}_p$
parameters.
\begin{equation}
\tilde{g}\rightarrow q\bar{q}\nu, ~~q\bar{q}^{\prime}l, ~~g\nu
\end{equation}
If gluino is the LSP and $\lambda^{\prime}$ is not extremely
small, the above ${\rlap/R}_p$ processes could be the only
possible decay channels and the gluino will decay inside the
detector. If the couplings are very small, the gluino can form
bound states with gluon or quarks before decaying, known as
R-hardrons\cite{gLSP}\cite{mglimit}.
\par
In Fig.4 we depict the dependence of the cross section for the
process $pp\rightarrow \tilde{g}\tau ^{+}(\tilde{g}\nu _\tau )+X$
on mass of gluino. All the cross sections of the single gluino
production via Drell-Yan and gluon fusion subprocesses are
plotted. In order to obtain a wide variety of gluino mass, we
abolish the MSUGRA model and choose $m_{\tilde{g}}$ and
$m_{\tilde{q}_{L,R}}$ arbitrarily. We consider a gluino with its
mass varying from 5 GeV, when gluino may be the LSP, up to 800
GeV. We assume that there is no mass mixing between $\tilde{q}_L$
and $\tilde{q}_R$ for all the up-type- and down-type-squarks of
the first two generations, since in general the mixing size is
proportional to the mass of the related ordinary
quark\cite{Ellis}. As a representation example of the parameter
space, we choose the squark masses as below,
$$
m_{\tilde{u}_L}=m_{\tilde{c}_L}=392.9~GeV,~
m_{\tilde{u}_R}=m_{\tilde{c}_R}=384.3~GeV,
$$
$$
m_{\tilde{d}_L}=m_{\tilde{s}_L}=400.0~GeV,~
m_{\tilde{d}_R}=m_{\tilde{s}_R}=385.1~GeV,
$$
$$
m_{\tilde{b}_1}=358.5~GeV,~m_{\tilde{b}_2}=385.0~GeV,~
m_{\tilde{t}_1}=312.5~GeV,~m_{\tilde{t}_2}=404.4~GeV.
$$
The total cross section of single gluino associated production
drops from 153 fb to 4 fb with the increment of $m_{\tilde{g}}$.
The figure shows that the single gluino production rate via
one-loop process $pp \to gg \to \tilde{g}+X$ is comparable with
those from other production mechanisms. We can see when the gluino
mass is less than $30~GeV$, the cross section of $pp \to gg \to
\tilde{g}+X$ can be over two times larger than the cross section
of $pp \to u\bar{d}(\bar{u}d, ~d\bar{d} ) \to \tilde{g}+X$
quantitatively. When the gluino mass is greater than $200~GeV$,
the contribution of gluon fusion to the gluino production process
drops to small values. If the integrated luminosity at the LHC is
30 $fb^{-1}$, typically $10^2\sim 10^3$ raw events can be produced
when we take the values of the ${\rlap/R}_p$ parameters are close
to the present upper bounds. With the parameters taken in Fig.4,
the signatures of gluinos can be detected in following ways:
\par
(1) If $m_{\tilde{g}}$ is heavier than $400~GeV$, the signals of
gluinos can be two, four or six jets together with energy missing,
which are induced by the decay of (17) and the subsequent cascade
decays of squarks (shown in Eqs.(18)and (19)). The number of jets
depends on the kinematical phase space of the decays (17), (18)
and (19)
\par
(2) If the mass of gluino is less than about $400~GeV$, the
gluinos have the possible decays shown in Eq.(20) and charginos
and neutralinos subsequent decays as in Eqs.(19). The gluino
signals are two or four jets with missing energy, and one or three
jets with a energetic lepton.
\par
(3)If the gluinos are the LSP or gluinos are almost degenerate
with the lightest neutralino and chargino, ${\rlap/R}_p$ decays of
gluinos shown in Eq.(21)will be significant, their signatures will
be one or two jets with missing energy, or two jets with an
energetic lepton.
\par
In Fig.5 we plotted the cross section in the MSUGRA scenario, with
$m_0$ varying from 100 GeV to 800 GeV. The other input parameters
are chosen as: $m_{\frac{1}{2}}=150~GeV,~A_0=300~GeV,~\tan
\beta=4$ and set $sign(\mu)=+$. With above MSUGRA input
parameters, we get the mass of gluino ranging from 395 GeV to 439
GeV, the lighter scalar top and bottom quarks
($\tilde{t}_1,\tilde{b}_1$) have the masses between 270 GeV to 493
GeV and 332 GeV to 700 GeV respectively, the first two generation
squarks are ranged from 346 GeV to 865 GeV. The calculation shows
that the gluino mass is not sensitive to $m_0$ with those input
parameters, then the production cross sections decrease gently
with the increment of $m_0$. In this case the contribution from
the one-loop process of $pp \to gg \to \tilde{g}+X$ reaches $1\%$
of the total cross section of single gluino associated production
at the LHC. As we have mentioned, the cross section of the gluon
fusion process is still not negligible since it depends on
different ${\rlap/R}_p$ parameters with the Drell-Yan process.
With these MSUGRA parameters, the gluino decays $\tilde{g}
\rightarrow q\tilde{q}, ~(q=u,d,c,s)$ are allowed kinematically
when $m_{0}$ is about 100 GeV, but the decays of Eq. (20) may be
used as gluino signatures when $m_{0}$ approaches 800 GeV.

\vskip 10mm \noindent
{\Large {\bf IV.Summary}} \vskip 5mm
\par
As shown in reference \cite{rgdecay}\cite{yu}, the gluino decay
can be detectable at the LHC even when R-parity is not conserved.
In this work we have studied the single gluino production
associated with a charged lepton or neutrino in the
${\rlap/{R}}_p$ MSSM at the LHC through the process $pp\rightarrow
\tilde{g}\tau ^{+}(\tilde{g}\nu _\tau )+X$. We investigated
contributions from both the tree-level Drell-Yan process and the
one-loop gluon fusion process. In the subprocesses
$d\bar{d}\rightarrow \tilde{g}\nu$ and $gg\rightarrow
\tilde{g}\nu$, a neutrino will be produced, which leads to an
energy missing by neutrino. We studied the dependence of the cross
section on $m_{\tilde{g}}$ and $m_0$. The results show that when
the $R_p$-violating coupling parameters are not too small and the
$m_{\tilde{g}}\sim 200~GeV$, the production rate of the single
gluino associated with a charged lepton or neutrino can reach 30
femto barn. That means $10^3$ raw events can be obtained at the
LHC with integrated luminosity $30~fb^{-1}$. We conclude that the
single gluino production associated with a charged lepton or
neutrino could be observable at the LHC, when the $R_p$-violating
parameters are close to the upper bounds. In this case we see the
single gluino production associated with a charged lepton or
neutrino can provide a quite prospective way to probe
supersymmetry and $R_p$ violation. Even if we couldn't find the
signal of single gluino production in the experiment, we can get
more stringent constraints on ${\rlap/R}_p$ couplings.

\vskip 5mm
\noindent{\large {\bf Acknowledgement:}}
\par
This work was supported in part by the National Natural Science Foundation
of China(project numbers: 19675033,10005009), the Education Ministry of
China and the State Commission of Science and Technology of China.

\vskip 5mm
\noindent{\large {\bf Appendix}}

\par
The Feynman rules for the R-parity violating couplings we used are
listed below:
\begin{eqnarray*}
&&u_I-l_K-\tilde{D}_{J,m}^c:-V_{u_Il_K\tilde{D}_{J,m}}^{(1)}C^{-1}P_L \\
&&d_I^c-l_K-\tilde{U}_{J,m}:V_{d_Il_K\tilde{U}_{J,m}}^{(1)}P_L \\
&&d_I-\nu _K-\tilde{D}_{J,m}^c:C^{-1}\left[ V_{d_I\nu _K\tilde{D}%
_{J,m}}^{(2)*}P_L+V_{d_I\nu _K\tilde{D}_{J,m}}^{(1)*}P_R\right]  \\
&&d_I^c-\nu _K-\tilde{D}_{J,m}:V_{d_I\nu _K\tilde{D}_{J,m}}^{(1)}P_L+V_{d_I%
\nu _K\tilde{D}_{J,m}}^{(2)}P_R
\end{eqnarray*}
where C is the charge conjugation operator, $P_{L,R}=\frac 12(1\mp
\gamma _5)$. The coefficients of vertices can be written as:
\begin{eqnarray*}
&&V_{u_Il_K\tilde{D}_{J,m}}^{(1)}=i\lambda _{KIJ}^{\prime
}Z_{D_J}^{2m*},V_{d_Il_K\tilde{U}_{J,m}}^{(1)}=i\lambda
_{KIJ}^{\prime
}Z_{U_J}^{1m*} \\
&&V_{d_I\nu _K\tilde{D}_{J,m}}^{(1)}=-i\lambda _{KJI}^{\prime
}Z_{D_J}^{1m*},V_{d_I\nu _K\tilde{D}_{J,m}}^{(2)}=-i\lambda
_{KIJ}^{\prime }Z_{D_J}^{2m}
\end{eqnarray*}
$Z_D^{ij}$ and $Z_U^{ij}$ are the matrices to diagonalize the
down-squark and up-squark mass matrices, respectively. We write
down form factors for one-loop diagrams of
the subprocess $g(p_1,a,\mu )g(p_2,b,\nu )\rightarrow \tilde{g}%
(k_1,c)\nu _\tau (k_2)$. The amplitude parts for the
$\hat{u}$-channel box diagrams can be obtained from the
$\hat{t}$-channel's by doing the exchanges as below:
\[
{\cal M}^{\hat{u}}={\cal M}^{\hat{t}}(\hat{t}\rightarrow
\hat{u},k_1\leftrightarrow k_2,\mu \leftrightarrow \nu
,a\leftrightarrow b)
\]
Then we present only the t-channel form factors for box diagrams.
In this appendix, we use the notifications defined below for
abbreviation:
\begin{eqnarray*}
&&C_0^{1,k},C_{ij}^{1,k}=C_0,C_{ij}[k_1,-p_1-p_2,m_{\tilde{d}_k},m_d,m_d] \\
&&C_0^{2,k},C_{ij}^{2,k}=C_0,C_{ij}[k_1,-p_1-p_2,m_d,m_{\tilde{d}_k},m_{%
\tilde{d}_k}] \\
&&C_0^{3,k},C_{ij}^{3,k}=C_0,C_{ij}[k_2,k_1,m_{\tilde{d}_k},m_d,m_{\tilde{d}%
_k}] \\
&&C_0^4=C_0[-p_1,-p_2,m_d,m_d,m_d] \\
&&D_0^{1,k},D_{ij}^{1,k},D_{ijl}^{1,k}=D_0,D_{ij},D_{ijl}[k_1,-p_1,-p_2,m_{%
\tilde{d}_k},m_d,m_d,m_d] \\
&&D_0^{2,k},D_{ij}^{2,k},D_{ijl}^{2,k}=D_0,D_{ij},D_{ijl}[k_1,-p_2,-p_1,m_d,m_{%
\tilde{d}_k},m_{\tilde{d}_k},m_{\tilde{d}_k}] \\
&&D_0^{3,k},D_{ij}^{3,k},D_{ijl}^{3,k}=D_0,D_{ij},D_{ijl}[-p_2,k_1,-p_1,m_{%
\tilde{d}_k},m_{\tilde{d}_k},m_d,m_d] \\
&&F_{1,i}^{\pm }=Z_D^{1i*}V_{D_id\nu }^{(1)}\pm
Z_D^{2i*}V_{D_id\nu }^{(2)}
\\
&&F_{2,i}^{\pm }=Z_D^{1i*}V_{D_id\nu }^{(2)}\pm
Z_D^{2i*}V_{D_id\nu }^{(1)}
\end{eqnarray*}
\par
The form factors in the amplitude of the quartic interaction
diagrams Fig.3(b) are expressed as
\begin{eqnarray*}
f_1^q &=& -\frac{g_s^3}{32\sqrt{2}\pi ^2}\sum%
\limits_{i=1}^2(C_0^{3,i}F_{2,i}^{-}m_d-C_{12}^{3,i}m_{\tilde{g}%
}F_{1,i}^{-})d_{abc}-h.c. \\
f_3^q &=& -\frac{g_s^3}{32\sqrt{2}\pi ^2}\sum%
\limits_{i=1}^2(C_0^{3,i}F_{2,i}^{+}m_d+C_{12}^{3,i}m_{\tilde{g}%
}F_{1,i}^{+})d_{abc}+h.c.
\end{eqnarray*}
\par
For the other form factors of the quartic interaction diagrams,
$f_i^q=0$. The none-zero form factors in the amplitude from the
triangle diagrams depicted in Fig.3(c) are listed below:
\begin{eqnarray*}
f_1^{tr} &=& \frac{i g_s^3 f_{abc}}{32\sqrt{2}\pi^2\hat{s}}
\sum\limits_{i=1}^2 \left\{ C_{21}^{1,i} m_{\tilde{g}} (\hat{t}-
\hat{u}) F_{1,i}^--C_0^{1,i} (\hat{s}-\hat{t}+\hat{u})
(m_{\tilde{g}} F_{1,i}^-+F_{2,i}^- m_d)-C_{11}^{1,i} \left[
m_{\tilde{g}} (\hat{s} \right.
\right. \\
&-& \left. 2 \hat{t}+2 \hat{u}) F_{1,i}^-+(-\hat{t}+\hat{u})
F_{2,i}^-m_d \right]+(\hat{t}-\hat{u}) \left[ C_{21}^{2,i}
m_{\tilde{g}}
F_{1,i}^--C_0^{2,i} F_{2,i}^- m_d \right. \\
&+& \left. \left. C_{11}^{2,i} (m_{\tilde{g}} F_{1,i}^-- m_d
F_{2,i}^-)
\right] \right\}-h.c. \\
f_3^{tr} &=& -\frac{i g_s^3 f_{abc}}{32\sqrt{2}\pi^2\hat{s}}
\sum\limits_{i=1}^2 \left\{ C_{21}^{1,i} m_{\tilde{g}} (\hat{t}-
\hat{u}) F_{1,i}^+-C_0^{1,i} (\hat{s}-\hat{t}+\hat{u})
(m_{\tilde{g}} F_{1,i}^+-F_{2,i}^+ m_d)-C_{11}^{1,i} \left[
m_{\tilde{g}} (\hat{s} \right.
\right. \\
&-& \left. 2 \hat{t}+2 \hat{u}) F_{1,i}^++(\hat{t}-\hat{u})
F_{2,i}^+ m_d \right]+(\hat{t}-\hat{u}) \left[ C_{21}^{2,i}
m_{\tilde{g}}
F_{1,i}^++C_0^{2,i} F_{2,i}^+ m_d \right. \\
&+& \left. \left.C_{11}^{2,i} (m_{\tilde{g}} F_{1,i}^++ m_d
F_{2,i}^+)
\right] \right\}+h.c. \\
f_7^{tr} &=& \frac{i g_s^3 f_{abc}}{32\sqrt{2}\pi^2\hat{s}}
\sum\limits_{i=1}^2 F_{1,i}^- \left\{ C_{22}^{1,i} \hat{s}+C_{21}^{1,i} m_{%
\tilde{g}}^2+C_{12}^{1,i} \left[ \hat{s}-2
(m_{\tilde{g}}^2-\hat{u}) \right]
-2 C_{24}^{1,i}+2 C_{11}^{1,i} m_{\tilde{g}}^2 \right. \\
&-& \left. 2 C_{23}^{1,i} (m_{\tilde{g}}^2-\hat{u}) +C_0^{1,i}(m_{\tilde{g}%
}^2-m_d^2)+2 C_{24}^{2,i}-(C_{12}^{2,i}+C_{23}^{2,i} )
(\hat{t}-\hat{u})
\right\}-h.c. \\
f_9^{tr} &=& -\frac{i g_s^3 f_{abc}}{32\sqrt{2}\pi^2\hat{s}}
\sum\limits_{i=1}^2 F_{1,i}^- \left\{C_{22}^{1,i} \hat{s}+C_{21}^{1,i} m_{%
\tilde{g}}^2+C_{12}^{1,i} \left[ \hat{s}-2
(m_{\tilde{g}}^2-\hat{t}) \right]
-2 C_{24}^{1,i}+2 C_{11}^{1,i} m_{\tilde{g}}^2 \right. \\
&-& \left. 2 C_{23}^{1,i} (m_{\tilde{g}}^2-\hat{t})+C_0^{1,i} (m_{\tilde{g}%
}^2-m_d^2)+2 C_{24}^{2,i}+(C_{12}^{2,i}+C_{23}^{2,i})
(\hat{t}-\hat{u})
\right\}-h.c. \\
f_{13}^{tr} &=& \frac{i g_s^3 f_{abc}}{32\sqrt{2}\pi^2\hat{s}}
\sum\limits_{i=1}^2 F_{1,i}^+ \left\{ C_{22}^{1,i} \hat{s}+C_{21}^{1,i} m_{%
\tilde{g}}^2+C_{12}^{1,i} \left[ \hat{s}-2
(m_{\tilde{g}}^2-\hat{u}) \right]
-2 C_{24}^{1,i}+2 C_{11}^{1,i} m_{\tilde{g}}^2 \right. \\
&-& \left. 2 C_{23}^{1,i} (m_{\tilde{g}}^2-\hat{u})+C_0^{1,i} (m_{\tilde{g}%
}^2-m_d^2) +2 C_{24}^{2,i}-(C_{12}^{2,i}+C_{23}^{2,i} )
(\hat{t}-\hat{u})
\right\}+h.c. \\
f_{15}^{tr} &=& -\frac{i g_s^3 f_{abc}}{32\sqrt{2}\pi^2\hat{s}}
\sum\limits_{i=1}^2 F_{1,i}^+ \left\{ C_{22}^{1,i} \hat{s}+C_{21}^{1,i} m_{%
\tilde{g}}^2+C_{12}^{1,i} \left[ \hat{s}-2
(m_{\tilde{g}}^2-\hat{t}) \right]
-2 C_{24}^{1,i}+2 C_{11}^{1,i} m_{\tilde{g}}^2 \right. \\
&-& \left. 2 C_{23}^{1,i} (m_{\tilde{g}}^2-\hat{t})+C_0^{1,i} (m_{\tilde{g}%
}^2-m_d^2)+2 C_{24}^{2,i}+(C_{12}^{2,i}+C_{23}^{2,i})
(\hat{t}-\hat{u})
\right\}+h.c. \\
f_{33}^{tr} &=& \frac{i g_s^3 f_{abc}}{16\sqrt{2}\pi^2\hat{s}}
\sum\limits_{i=1}^2 \left[ C_{11}^{1,i} m_{\tilde{g}}
F_{1,i}^-+C_0^{1,i}
(m_{\tilde{g}} F_{1,i}^-+ m_d F_{2,i}^-) \right]-h.c. \\
f_{34}^{tr} &=& -\frac{i g_s^3 f_{abc}}{16\sqrt{2}\pi^2\hat{s}}
\sum\limits_{i=1}^2 \left[ C_{11}^{1,i} m_{\tilde{g}}
F_{1,i}^++C_0^{1,i} (m_{\tilde{g}} F_{1,i}^+- m_d F_{2,i}^+)
\right]+h.c.
\end{eqnarray*}
\par
The none-zero form factors of the amplitude part form t-channel
box diagrams, Fig.3(a) are written as
\begin{eqnarray*}
f_1^{b,t} &=& -\frac{g_s^3}{16\sqrt{2}\pi^2} (d_{abc}-i f_{abc})
\left[ D_{311}^{2,i} m_{\tilde{g}} F_{1,i}^--D_{27}^{2,i}
F_{2,i}^-
m_d-D_{312}^{3,i} m_{\tilde{g}} F_{1,i}^- \right. \\
&-& \left. D_{27}^{3,i} (m_{\tilde{g}} F_{1,i}^-+F_{2,i}^-
m_d)-D_{311}^{1,i} m_{\tilde{g}} F_{1,i}^-- D_{27}^{1,i}
(m_{\tilde{g}}
F_{1,i}^-+m_d F_{2,i}^-) \right]-h.c. \\
f_2^{b,t} &=& -\frac{g_s^3}{16\sqrt{2}\pi^2} (d_{abc}-i f_{abc})
\left[ -D_{31}^{2,i} m_{\tilde{g}} F_{1,i}^-+ D_0^{2,i} F_{2,i}^-
m_d-D_{11}^{2,i}
(m_{\tilde{g}} F_{1,i}^--2 F_{2,i}^- m_d) \right. \\
&-& \left. D_{12}^{2,i} (2 m_{\tilde{g}} F_{1,i}^--F_{2,i}^-
m_d)+D_{32}^{3,i} m_{\tilde{g}} F_{1,i}^-+ D_{12}^{3,i}
(m_{\tilde{g}}
F_{1,i}^-+F_{2,i}^- m_d) \right. \\
&+& \left.D_{22}^{3,i} (2 m_{\tilde{g}} F_{1,i}^-+F_{2,i}^-
m_d)+D_{31}^{1,i} m_{\tilde{g}} F_{1,i}^-+D_0^{1,i} (m_{\tilde{g}}
F_{1,i}^-+F_{2,i}^- m_d) \right. \\
&+& \left. D_{21}^{1,i} (3 m_{\tilde{g}} F_{1,i}^-+F_{2,i}^- m_d)
+D_{11}^{1,i} (3 m_{\tilde{g}} F_{1,i}^-+2 F_{2,i}^- m_d) \right]-h.c. \\
f_3^{b,t} &=& \frac{g_s^3}{16\sqrt{2}\pi^2} (d_{abc}-i f_{abc})
\left[ D_{311}^{2,i} m_{\tilde{g}} F_{1,i}^++D_{27}^{2,i}
F_{2,i}^+
m_d-D_{312}^{3,i} m_{\tilde{g}} F_{1,i}^+ \right. \\
&-& \left. D_{27}^{3,i} (m_{\tilde{g}} F_{1,i}^+-F_{2,i}^+
m_d)-D_{311}^{1,i} m_{\tilde{g}} F_{1,i}^+- D_{27}^{1,i}
(m_{\tilde{g}}
F_{1,i}^+-F_{2,i}^+ m_d) \right]+h.c. \\
f_4^{b,t} &=& -\frac{g_s^3}{16\sqrt{2}\pi^2} (d_{abc}-i f_{abc})
\left[ D_{31}^{2,i} m_{\tilde{g}} F_{1,i}^++D_0^{2,i} F_{2,i}^+
m_d+D_{12}^{2,i} (2
m_{\tilde{g}} F_{1,i}^++F_{2,i}^+ m_d) \right. \\
&+& D_{11}^{2,i} (m_{\tilde{g}} F_{1,i}^++2 F_{2,i}^+ m_d)-D_{32}^{3,i} m_{%
\tilde{g}} F_{1,i}^+- D_{12}^{3,i} (m_{\tilde{g}} F_{1,i}^+-F_{2,i}^+ m_d) \\
&-& \left. D_{22}^{3,i} (2 m_{\tilde{g}} F_{1,i}^+-F_{2,i}^+
m_d)-D_{31}^{1,i} m_{\tilde{g}} F_{1,i}^+- D_{11}^{1,i} (3
m_{\tilde{g}}
F_{1,i}^+-2 F_{2,i}^+ m_d) \right. \\
&-& \left. D_0^{1,i} (m_{\tilde{g}} F_{1,i}^+-F_{2,i}^+
m_d)-D_{21}^{1,i} (3
m_{\tilde{g}} F_{1,i}^+-F_{2,i}^+ m_d) \right]+h.c. \\
f_5^{b,t} &=& -\frac{g_s^3}{32\sqrt{2}\pi^2} (d_{abc}-i f_{abc})
F_{1,i}^- \left[ 2 (D_{27}^{2,i}+D_{311}^{2,i})+2 D_{27}^{3,i}+4
D_{312}^{3,i}-
D_{32}^{3,i} m_{\tilde{g}}^2-D_{310}^{3,i} \hat{s} \right. \\
&+& D_{38}^{3,i} (m_{\tilde{g}}^2-\hat{t})+D_{26}^{3,i} (m_{\tilde{g}}^2-%
\hat{s}- \hat{t})+D_{36}^{3,i} (m_{\tilde{g}}^2-\hat{u})+D_{24}^{3,i} (m_{%
\tilde{g}}^2-\hat{s}- \hat{u}) \\
&-& \left. D_{22}^{3,i} (m_{\tilde{g}}^2+\hat{u})-D_{12}^{3,i} (\hat{s}+\hat{%
u}-m_d^2)+4 (D_{27}^{1,i}+D_{311}^{1,i})+(-3 D_{21}^{1,i}-D_{31}^{1,i}) m_{%
\tilde{g}}^2 - (D_{26}^{1,i}+D_{310}^{1,i}) \hat{s}\right. \\
&+& D_{12}^{1,i} (m_{\tilde{g}}^2-\hat{t})+2 D_{24}^{1,i} (m_{\tilde{g}}^2-%
\hat{t})+D_{34}^{1,i} (m_{\tilde{g}}^2-\hat{t})+ D_{25}^{1,i} (2 m_{\tilde{g}%
}^2-\hat{s}-2 \hat{u}) \\
&+& \left. D_{35}^{1,i} (m_{\tilde{g}}^2-\hat{u})+D_{13}^{1,i} (m_{\tilde{g}%
}^2- \hat{s}-\hat{u})-D_0^{1,i} (m_{\tilde{g}}^2-m_d^2)-D_{11}^{1,i} (3 m_{%
\tilde{g}}^2-m_d^2) \right]-h.c. \\
f_6^{b,t} &=& -\frac{g_s^3}{32\sqrt{2}\pi^2} (d_{abc}-i f_{abc})
F_{1,i}^- \left[ 2 D_{27}^{2,i}+2 D_{311}^{2,i}-2 D_{27}^{3,i}-2
D_{312}^{3,i}-C_0^4+2
D_{27}^{1,i} \right. \\
&-& \left. 2 D_{311}^{1,i}- D_{21}^{1,i}
m_{\tilde{g}}^2-D_{26}^{1,i} \hat{s}
+ D_{24}^{1,i} (m_{\tilde{g}}^2-\hat{t})+D_{25}^{1,i} (m_{\tilde{g}}^2-\hat{u%
})+D_0^{1,i} m_{\tilde{d}_i}^2 \right]-h.c. \\
f_7^{b,t} &=& -\frac{g_s^3}{16\sqrt{2}\pi^2} (d_{abc}-i f_{abc})
F_{1,i}^-
(-D_{313}^{2,i}+D_{27}^{3,i}+D_{313}^{3,i}+D_{27}^{1,i}+D_{312}^{1,i})-h.c.
\\
f_8^{b,t} &=& -\frac{g_s^3}{16\sqrt{2}\pi^2} (d_{abc}-i f_{abc})
F_{1,i}^- ( 2
D_{25}^{2,i}+D_{13}^{2,i}+D_{35}^{2,i}-D_{26}^{3,i}-D_{38}^{3,i}-2
D_{24}^{1,i}-D_{12}^{1,i}-D_{34}^{1,i} )-h.c. \\
f_9^{b,t} &=& -\frac{g_s^3}{16\sqrt{2}\pi^2} (d_{abc}-i f_{abc})
F_{1,i}^-
(-D_{312}^{2,i}+D_{27}^{3,i}+D_{311}^{3,i}+D_{313}^{1,i})-h.c. \\
f_{10}^{b,t} &=& -\frac{g_s^3}{16\sqrt{2}\pi^2} (d_{abc}-i
f_{abc}) F_{1,i}^- ( 2
D_{24}^{2,i}+D_{12}^{2,i}+D_{34}^{2,i}-D_{12}^{3,i}-D_{22}^{3,i}-D_{24}^{3,i}-
D_{36}^{3,i}-2 D_{25}^{1,i}
\\
&-& D_{13}^{1,i}-D_{35}^{1,i} )-h.c. \\
f_{11}^{b,t} &=& -\frac{g_s^3}{32\sqrt{2}\pi^2} (d_{abc}-i
f_{abc}) F_{1,i}^+ \left[ 2 (D_{27}^{2,i}+D_{311}^{2,i})+2
D_{27}^{3,i}+4 D_{312}^{3,i}-
D_{32}^{3,i} m_{\tilde{g}}^2-D_{310}^{3,i} \hat{s} \right. \\
&+& D_{38}^{3,i} (m_{\tilde{g}}^2-\hat{t})+D_{26}^{3,i} (m_{\tilde{g}}^2-%
\hat{s}- \hat{t})+D_{36}^{3,i} (m_{\tilde{g}}^2-\hat{u})+D_{24}^{3,i} (m_{%
\tilde{g}}^2-\hat{s}- \hat{u}) \\
&-& \left. D_{22}^{3,i} (m_{\tilde{g}}^2+\hat{u})-D_{12}^{3,i} (\hat{s}+\hat{%
u}-m_d^2)+4 (D_{27}^{1,i}+D_{311}^{1,i})+(-3 D_{21}^{1,i}-D_{31}^{1,i}) m_{%
\tilde{g}}^2 \right. \\
&-& (D_{26}^{1,i}+D_{310}^{1,i}) \hat{s}+D_{12}^{1,i} (m_{\tilde{g}}^2- \hat{%
t})+2 D_{24}^{1,i} (m_{\tilde{g}}^2-\hat{t})+D_{34}^{1,i} (m_{\tilde{g}}^2-%
\hat{t}) \\
&+& D_{25}^{1,i} (2 m_{\tilde{g}}^2-\hat{s}-2 \hat{u})+D_{35}^{1,i} (m_{%
\tilde{g}}^2-\hat{u})+D_{13}^{1,i} (m_{\tilde{g}}^2-
\hat{s}-\hat{u})
-D_0^{1,i} (m_{\tilde{g}}^2-m_d^2) \\
&-& \left. D_{11}^{1,i} (3 m_{\tilde{g}}^2-m_d^2) \right]+h.c. \\
f_{12}^{b,t} &=& -\frac{g_s^3}{32\sqrt{2}\pi^2} (d_{abc}-i
f_{abc}) F_{1,i}^+ \left[ 2 D_{27}^{2,i}+2 D_{311}^{2,i}-2
D_{27}^{3,i}-2 D_{312}^{3,i}-C_0^4+2
D_{27}^{1,i} \right. \\
&-& \left. 2 D_{311}^{1,i}- D_{21}^{1,i}
m_{\tilde{g}}^2-D_{26}^{1,i} \hat{s}
+D_{24}^{1,i} (m_{\tilde{g}}^2-\hat{t})+D_{25}^{1,i} (m_{\tilde{g}}^2-\hat{u}%
)+D_0^{1,i} m_{\tilde{d}_i}^2 \right]+h.c. \\
f_{13}^{b,t} &=& -\frac{g_s^3}{16\sqrt{2}\pi^2} (d_{abc}-i
f_{abc}) F_{1,i}^+
(-D_{313}^{2,i}+D_{27}^{3,i}+D_{313}^{3,i}+D_{27}^{1,i}+D_{312}^{1,i})+h.c.
\\
f_{14}^{b,t} &=& -\frac{g_s^3}{16\sqrt{2}\pi^2} (d_{abc}-i
f_{abc}) F_{1,i}^+ ( 2
D_{25}^{2,i}+D_{13}^{2,i}+D_{35}^{2,i}-D_{26}^{3,i}-D_{38}^{3,i}-2
D_{24}^{1,i}-D_{12}^{1,i}-D_{34}^{1,i} )+h.c. \\
f_{15}^{b,t} &=& -\frac{g_s^3}{16\sqrt{2}\pi^2} (d_{abc}-i
f_{abc}) F_{1,i}^+
(-D_{312}^{2,i}+D_{27}^{3,i}+D_{311}^{3,i}+D_{313}^{1,i})+h.c. \\
f_{16}^{b,t} &=& -\frac{g_s^3}{16\sqrt{2}\pi^2} (d_{abc}-i
f_{abc}) F_{1,i}^+ ( 2
D_{24}^{2,i}+D_{12}^{2,i}+D_{34}^{2,i}-D_{12}^{3,i}-D_{22}^{3,i}-D_{24}^{3,i}
-D_{36}^{3,i} -2 D_{25}^{1,i}
\\
&-& D_{13}^{1,i}-D_{35}^{1,i} )+h.c. \\
f_{17}^{b,t} &=& -\frac{g_s^3}{64\sqrt{2}\pi^2} (d_{abc}-i
f_{abc}) \left\{\left[-D_{31}^{1,i} m_{\tilde{g}}^2+ 6
D_{311}^{1,i}-D_{310}^{1,i}
\hat{s}+D_{34}^{1,i} (m_{\tilde{g}}^2-\hat{t})+D_{35}^{1,i} (m_{\tilde{g}%
}^2- \hat{u}) \right] m_{\tilde{g}} F_{1,i}^- \right. \\
&+& 2 D_{27}^{1,i} (3 m_{\tilde{g}} F_{1,i}^-+2 F_{2,i}^- m_d)+
D_{24}^{1,i}
(m_{\tilde{g}}^2-\hat{t}) (2 m_{\tilde{g}} F_{1,i}^-+F_{2,i}^- m_d) \\
&-& D_{21}^{1,i} m_{\tilde{g}} \left[ (2 m_{\tilde{g}}^2+\hat{t})
F_{1,i}^-+
m_{\tilde{g}} F_{2,i}^- m_d \right]+D_{25}^{1,i} \left[ m_{\tilde{g}} (2 m_{%
\tilde{g}}^2-\hat{s}-2 \hat{u})
F_{1,i}^-+(m_{\tilde{g}}^2-\hat{u})
F_{2,i}^- m_d \right] \\
&-& D_{11}^{1,i} \left[ (m_{\tilde{g}}^2+\hat{t}) F_{2,i}^- m_d+m_{\tilde{g}%
} F_{1,i}^- (m_{\tilde{g}}^2+2 \hat{t}-m_d^2) \right] \\
&+& \left. (m_{\tilde{g}} F_{1,i}^-+F_{2,i}^- m_d) \left[ -D_{26}^{1,i} \hat{%
s}+D_{12}^{1,i} (m_{\tilde{g}}^2-\hat{t})+D_{13}^{1,i} ( m_{\tilde{g}}^2-%
\hat{s}-\hat{u})+D_0^{1,i} (-\hat{t}+m_d^2) \right] \right\}-h.c. \\
f_{18}^{b,t} &=& \frac{g_s^3}{32\sqrt{2}\pi^2} (d_{abc}-i f_{abc})
\left[ D_{22}^{3,i} m_{\tilde{g}} F_{1,i}^-+D_{12}^{3,i}
(m_{\tilde{g}}
F_{1,i}^-+F_{2,i}^- m_d)+D_{21}^{1,i} m_{\tilde{g}} F_{1,i}^- \right. \\
&+& \left. D_0^{1,i} (m_{\tilde{g}} F_{1,i}^-+F_{2,i}^-
m_d)+D_{11}^{1,i} (2
m_{\tilde{g}} F_{1,i}^-+F_{2,i}^- m_d) \right]-h.c. \\
f_{19}^{b,t} &=& \frac{g_s^3}{32\sqrt{2}\pi^2} (d_{abc}-i f_{abc})
\left[ D_{21}^{1,i} m_{\tilde{g}} F_{1,i}^-+D_0^{1,i}
(m_{\tilde{g}} F_{1,i}^-+F_{2,i}^- m_d)+D_{11}^{1,i} (2
m_{\tilde{g}} F_{1,i}^-+F_{2,i}^-
m_d) \right]-h.c. \\
f_{20}^{b,t} &=& -\frac{g_s^3}{64\sqrt{2}\pi^2} (d_{abc}-i
f_{abc}) \left\{
\left[ D_{31}^{1,i} m_{\tilde{g}}^2-6 D_{311}^{1,i}+D_{310}^{1,i} \hat{s}%
-D_{34}^{1,i} (m_{\tilde{g}}^2-\hat{t})-D_{35}^{1,i} (m_{\tilde{g}}^2-\hat{u}%
) \right] m_{\tilde{g}} F_{1,i}^+ \right. \\
&-& 2 D_{27}^{1,i} (3 m_{\tilde{g}} F_{1,i}^+-2 F_{2,i}^+
m_d)-D_{24}^{1,i}
(m_{\tilde{g}}^2-\hat{t}) (2 m_{\tilde{g}} F_{1,i}^+-F_{2,i}^+ m_d) \\
&+& D_{21}^{1,i} m_{\tilde{g}} \left[ (2 m_{\tilde{g}}^2+\hat{t})
F_{1,i}^+-
m_{\tilde{g}} F_{2,i}^+ m_d \right]-D_{25}^{1,i} \left[ m_{\tilde{g}} (2 m_{%
\tilde{g}}^2-\hat{s}-2 \hat{u})
F_{1,i}^++(-m_{\tilde{g}}^2+\hat{u})
F_{2,i}^+ m_d \right] \\
&+& (m_{\tilde{g}} F_{1,i}^+-F_{2,i}^+ m_d) \left[ D_{26}^{1,i} \hat{s}%
-D_{12}^{1,i} (m_{\tilde{g}}^2-\hat{t})-D_{13}^{1,i} (m_{\tilde{g}}^2-\hat{s}%
-\hat{u}) +D_0^{1,i} (\hat{t}-m_d^2) \right] \\
&+& \left. D_{11}^{1,i} \left[ -(m_{\tilde{g}}^2+\hat{t}) F_{2,i}^+ m_d+ m_{%
\tilde{g}} F_{1,i}^+ (m_{\tilde{g}}^2+2 \hat{t}-m_d^2) \right]
\right\}+h.c.
\\
f_{21}^{b,t} &=& -\frac{g_s^3}{32\sqrt{2}\pi^2} (d_{abc}-i
f_{abc}) \left[ D_{22}^{3,i} m_{\tilde{g}} F_{1,i}^++ D_{12}^{3,i}
(m_{\tilde{g}}
F_{1,i}^+-F_{2,i}^+ m_d)+D_{21}^{1,i} m_{\tilde{g}} F_{1,i}^+ \right. \\
&+& \left. D_0^{1,i} ( m_{\tilde{g}} F_{1,i}^+-F_{2,i}^+
m_d)+D_{11}^{1,i}
(2 m_{\tilde{g}} F_{1,i}^+-F_{2,i}^+ m_d) \right]+h.c. \\
f_{22}^{b,t} &=& -\frac{g_s^3}{32\sqrt{2}\pi^2} (d_{abc}-i
f_{abc}) \left[ D_{21}^{1,i} m_{\tilde{g}} F_{1,i}^++D_0^{1,i} (
m_{\tilde{g}} F_{1,i}^+-F_{2,i}^+ m_d)+D_{11}^{1,i} (2
m_{\tilde{g}} F_{1,i}^+-F_{2,i}^+
m_d) \right]+h.c. \\
f_{23}^{b,t}&=& -\frac{g_s^3}{64\sqrt{2}\pi^2} (d_{abc}-i f_{abc})
F_{1,i}^-
\left[ -2 D_{27}^{3,i}-4 D_{27}^{1,i}-6 D_{312}^{1,i}+D_{34}^{1,i} m_{\tilde{%
g}}^2+D_{38}^{1,i} \hat{s}+(D_{22}^{1,i}+D_{36}^{1,i}) (-m_{\tilde{g}}^2+%
\hat{t}) \right. \\
&+& \left. D_{24}^{1,i} (m_{\tilde{g}}^2+ \hat{t})-D_{310}^{1,i} (m_{\tilde{g%
}}^2-\hat{u})-D_{26}^{1,i} (m_{\tilde{g}}^2-\hat{s}- \hat{u})+D_{12}^{1,i} (%
\hat{t}-m_d^2) \right]-h.c. \\
f_{24}^{b,t} &=& -\frac{g_s^3}{64\sqrt{2}\pi^2} (d_{abc}-i
f_{abc}) F_{1,i}^- \left[ -2 D_{27}^{1,i}-6 D_{313}^{1,i}+( 2
D_{11}^{1,i}+D_{21}^{1,i}+D_{35}^{1,i}) m_{\tilde{g}}^2 \right. \\
&+& D_{39}^{1,i} \hat{s}-D_{310}^{1,i} (m_{\tilde{g}}^2-\hat{t})
-D_{26}^{1,i}
(m_{\tilde{g}}^2-\hat{s}-\hat{t})+(D_{12}^{1,i}+D_{24}^{1,i})
(-m_{\tilde{g}}^2+\hat{t})-D_{37}^{1,i} ( m_{\tilde{g}}^2-\hat{u}) \\
&-& \left. D_{23}^{1,i} (m_{\tilde{g}}^2-\hat{s}-\hat{u})+D_{25}^{1,i} (\hat{%
t}+\hat{u})+D_0^{1,i} (m_{\tilde{g}}^2-m_d^2)-D_{13}^{1,i}
(m_{\tilde{g}}^2-
\hat{s}-\hat{t}-\hat{u}+m_d^2) \right]-h.c. \\
f_{25}^{b,t} &=& -\frac{g_s^3}{32\sqrt{2}\pi^2} (d_{abc}-i
f_{abc}) F_{1,i}^-
(D_{12}^{3,i}+D_{24}^{3,i}+D_{13}^{1,i}+D_{25}^{1,i})-h.c. \\
f_{26}^{b,t} &=& \frac{g_s^3}{32\sqrt{2}\pi^2} (d_{abc}-i f_{abc})
F_{1,i}^- (D_{12}^{1,i}+D_{24}^{1,i})-h.c. \\
f_{27}^{b,t} &=& -\frac{g_s^3}{64\sqrt{2}\pi^2} (d_{abc}-i
f_{abc}) F_{1,i}^+
\left[ -2 D_{27}^{3,i}-4 D_{27}^{1,i}-6 D_{312}^{1,i}+D_{34}^{1,i} m_{\tilde{%
g}}^2+D_{38}^{1,i} \hat{s}+(D_{22}^{1,i}+D_{36}^{1,i}) (-m_{\tilde{g}}^2+%
\hat{t}) \right. \\
&+& \left. D_{24}^{1,i} (m_{\tilde{g}}^2+ \hat{t})-D_{310}^{1,i} (m_{\tilde{g%
}}^2-\hat{u})-D_{26}^{1,i} (m_{\tilde{g}}^2-\hat{s}- \hat{u})+D_{12}^{1,i} (%
\hat{t}-m_d^2) \right]+h.c. \\
f_{28}^{b,t} &=& -\frac{g_s^3}{64\sqrt{2}\pi^2} (d_{abc}-i
f_{abc}) F_{1,i}^+ \left[ -2 D_{27}^{1,i}-6 D_{313}^{1,i}+(2
D_{11}^{1,i}+D_{21}^{1,i}+D_{35}^{1,i}) m_{\tilde{g}}^2 \right. \\
&+& D_{39}^{1,i} \hat{s}-D_{310}^{1,i} (m_{\tilde{g}}^2-\hat{t})
-D_{26}^{1,i}
(m_{\tilde{g}}^2-\hat{s}-\hat{t})+(D_{12}^{1,i}+D_{24}^{1,i})
(-m_{\tilde{g}}^2+\hat{t})-D_{37}^{1,i} ( m_{\tilde{g}}^2-\hat{u}) \\
&-& \left. D_{23}^{1,i} (m_{\tilde{g}}^2-\hat{s}-\hat{u})+D_{25}^{1,i} (\hat{%
t}+\hat{u})+D_0^{1,i} (m_{\tilde{g}}^2-m_d^2)-D_{13}^{1,i}
(m_{\tilde{g}}^2-
\hat{s}-\hat{t}-\hat{u}+m_d^2) \right]+h.c. \\
f_{29}^{b,t} &=& -\frac{g_s^3}{32\sqrt{2}\pi^2} (d_{abc}-i
f_{abc}) F_{1,i}^+
(D_{12}^{3,i}+D_{24}^{3,i}+D_{13}^{1,i}+D_{25}^{1,i})+h.c. \\
f_{30}^{b,t} &=& \frac{g_s^3}{32\sqrt{2}\pi^2} (d_{abc}- i
f_{abc})
F_{1,i}^+ (D_{12}^{1,i}+D_{24}^{1,i})+h.c. \\
f_{31}^{b,t} &=& \frac{g_s^3}{64\sqrt{2}\pi^2} (d_{abc}-i f_{abc})
\left[ D_{11}^{1,i} m_{\tilde{g}} F_{1,i}^-+D_0^{1,i}
(m_{\tilde{g}}
F_{1,i}^-+F_{2,i}^- m_d) \right]-h.c. \\
f_{32}^{b,t} &=& -\frac{g_s^3}{64\sqrt{2}\pi^2} (d_{abc}-i
f_{abc}) \left[ D_{11}^{1,i} m_{\tilde{g}} F_{1,i}^+ +D_0^{1,i}
(m_{\tilde{g}} F_{1,i}^+-F_{2,i}^+ m_d) \right]+h.c.
\end{eqnarray*}
\par
In this work we follow the definitions of two-, three-, and
four-point one loop integral functions of Passarino-Veltman as
shown in Ref.\cite{bernd}. All the vector and tensor integrals can
be calculated by deducing them into the forms of scalar integrals
\cite{veltman}.
\newline

\vskip 20mm 

\vskip 20mm \noindent{\Large {\bf Figure captions}} \vskip 5mm

\noindent

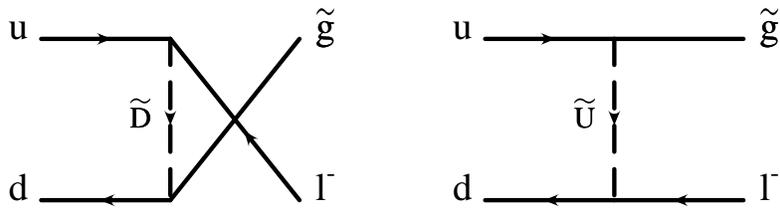
{\bf Fig.1} The ${\rlap/ R}_p$ MSSM tree-level diagrams of the process $u%
\bar{d}\rightarrow \tilde{g}l^+$.

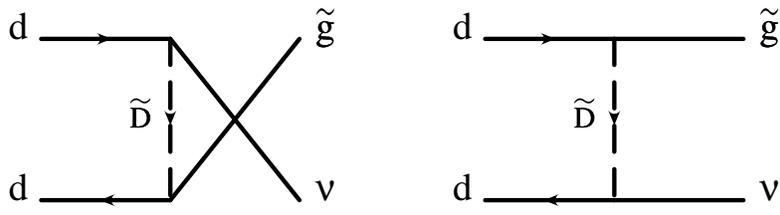
{\bf Fig.2} The ${\rlap/ R}_p$ MSSM tree-level diagrams of the process $d%
\bar{d}\rightarrow \tilde{g}\nu$.

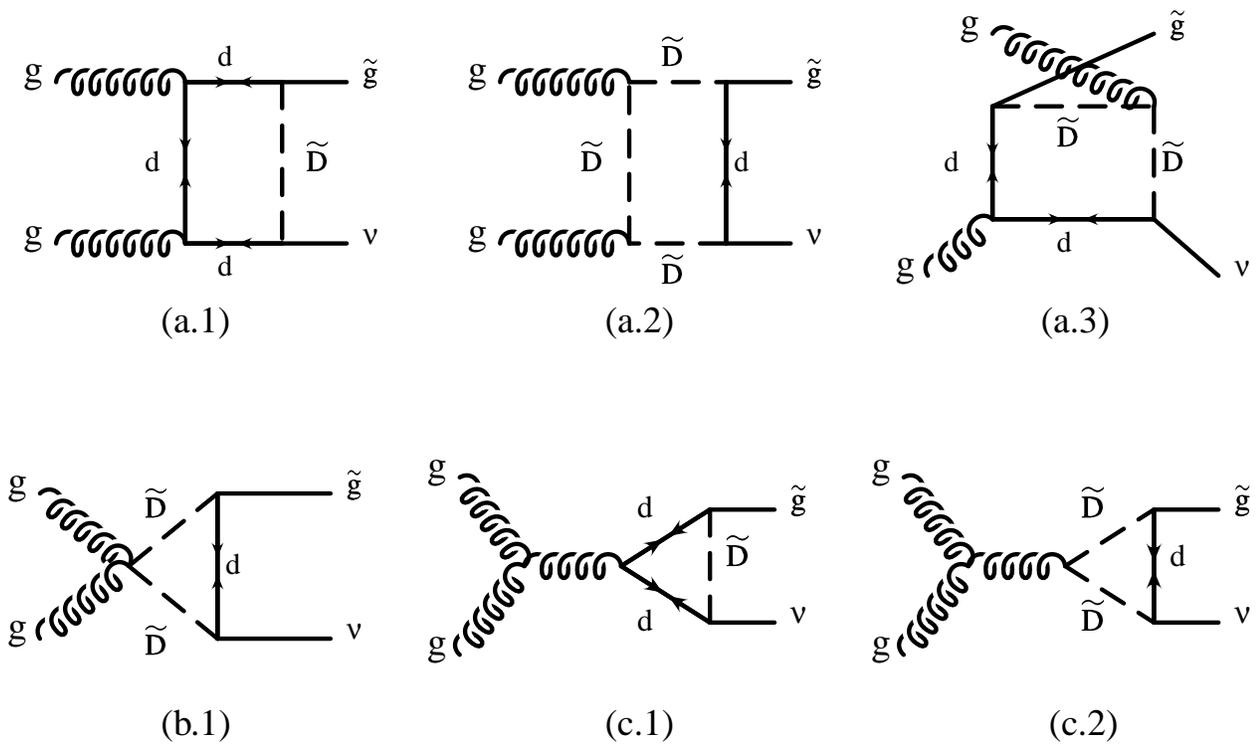
{\bf Fig.3} The ${\rlap/ R}_p$ MSSM one-loop diagrams of the process $%
gg\rightarrow\tilde{g}\nu$. (a) box diagrams; (b) quartic coupling
diagrams; (c) triangle diagrams.

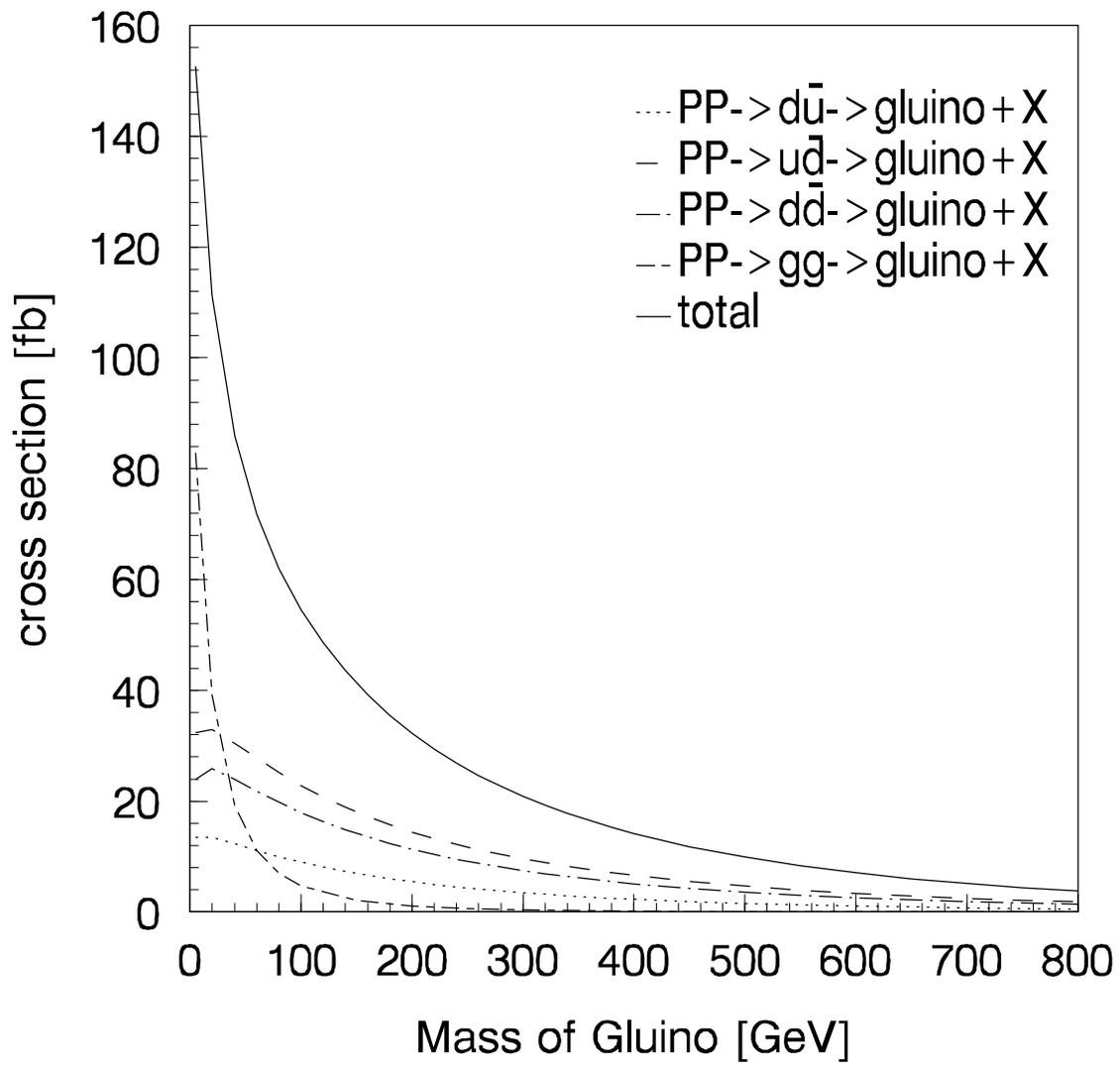
{\bf Fig.4} The cross sections of all the processes contributing
to the production of single gluino associated with a charged
lepton or neutrino at the LHC are depicted as functions of the
gluino mass, with the c.m. collision energy $\sqrt{s}$ at 14 TeV
and $m_{\tilde{g}}$ varying from 5 GeV to 800 GeV. The full-line
is for total cross section of single gluino production associated
with a charged lepton or neutrino $pp \to \tilde{g}+X$, The
dotted-line for $pp \to \bar{u}d \to \tilde{g}+X$. The dashed-line
for $pp \to u\bar{d} \to \tilde{g}+X$. The dash-dotted-line for
$pp \to d\bar{d} \to \tilde{g}+X$. The long-dashed and
short-dashed line for $pp \to gg \to \tilde{g}+X$.

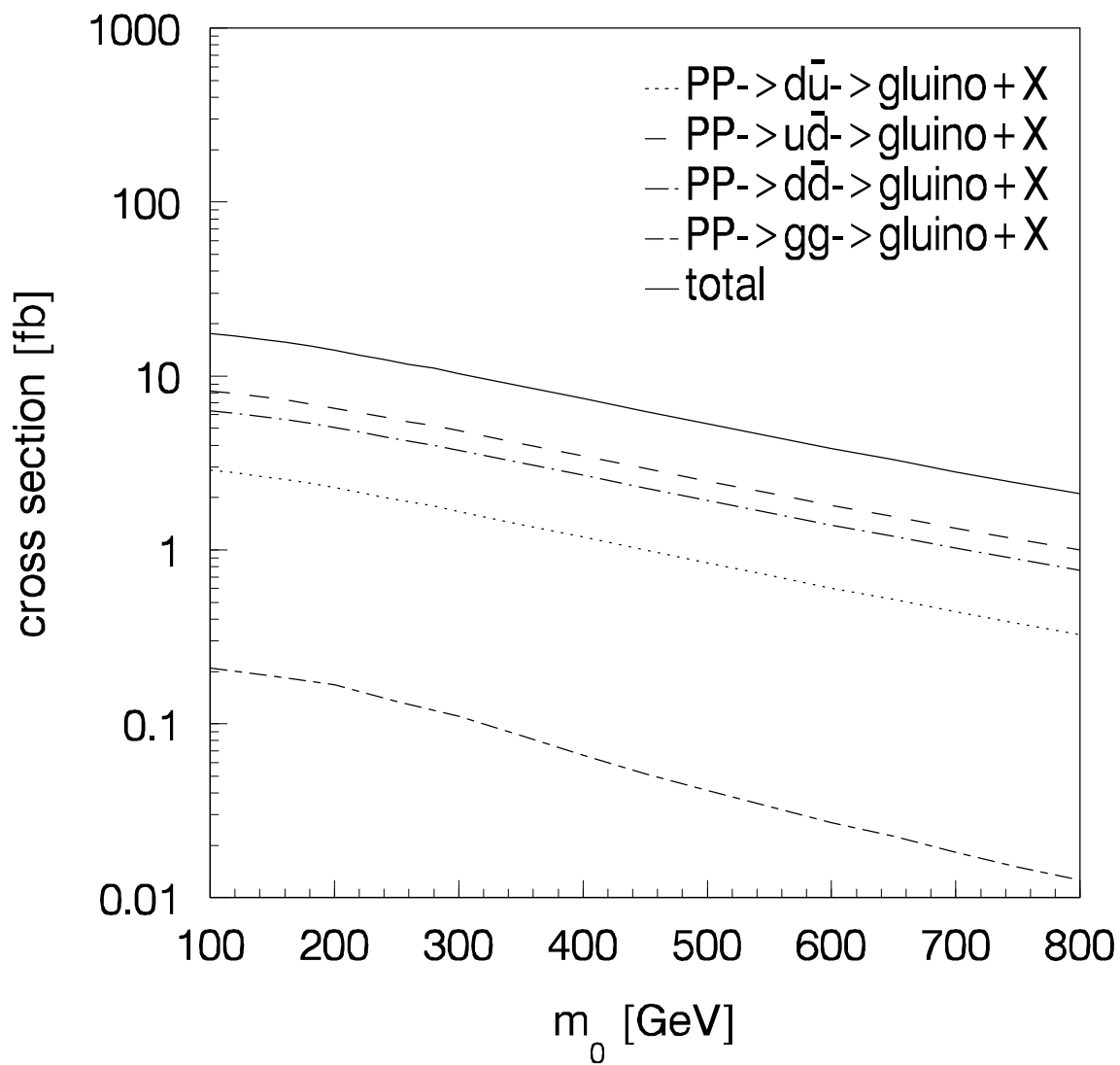
{\bf Fig.5} The cross sections of all the processes contributing
to the production of single gluino associated with a charged
lepton or neutrino at the LHC are depicted as functions of
parameter $m_0$ in the scenario of $R_p$ conserved MSUGRA, with
the c.m. collision energy $\sqrt{s}$ at 14 TeV. The parameter
$m_0$ is chosen to vary from 100 GeV to 800 GeV. The full-line is
for total cross section of single gluino production associated
with a charged lepton or neutrino $pp \to \tilde{g}+X$, The
dotted-line for $pp \to \bar{u}d \to \tilde{g}+X$. The dash-line
for $pp \to u\bar{d} \to \tilde{g}+X$. The dash-dotted-line for
$pp \to d\bar{d} \to \tilde{g}+X$. The long-dashed and
short-dashed line for $pp \to gg \to \tilde{g}+X$.

\vskip 3mm \noindent

\vskip 3mm 

\end{document}